\documentclass[preprint,12pt,authoryear]{elsarticle}

\usepackage{amssymb}
\usepackage{amsmath}
\usepackage{xcolor}

\journal{Astronomy and Computing}

\begin{document}

\begin{frontmatter}



\title{A general relativistic hydrodynamic simulation code for studying advective, sub-Keplerian accretion flow onto black holes} 


\author[l1]{Sudip K Garain} 
\ead{sgarain@iiserkol.ac.in}
\affiliation[l1]{organization={Department of Physical Sciences and Center of Excellence in Space Sciences India, Indian Institute of Science Education and Research Kolkata},
    addressline={Mohanpur}, 
    city={Nadia},
    postcode={741246}, 
    state={WB},
    country={India}}

\begin{abstract}
In this paper, we describe a general relativistic
hydrodynamics simulation code which is developed to simulate
advective accretion
flow onto black holes. We are particularly interested in
the accretion simulations of sub-Keplerian matter in the close vicinity
of black holes. Due to the presence of centrifugal barrier,
a nearly free-falling sub-Keplerian accretion flow slows down close to
a black hole and can even pass through shocks before
accelerating again to the black hole. We design our simulation
code using the high resolution shock capturing scheme so that
such shock structures can be captured and analyzed for relevance.
In this paper, we describe our implementation and validation
of the code against a few known analytical and numerical results
of sub-Keplerian matter accretion.
\end{abstract}




\end{frontmatter}



\section{Introduction}
\label{sec:1}
In a black hole X-ray binary (BHXRB) system or at the center 
of a galaxy, usually rotating fluid consisting of mostly 
plasma is accreted onto a black hole. Strong gravity in the close 
vicinity of the black hole affects fluid motion in a way 
that Newtonian potential cannot explain.
For example, the effective potential experienced by a rotating 
particle in the general relativistic calculation shows the presence of a
finite height potential barrier in contrast to an infinite barrier
as in the calculation using Newtonian gravitational field \citep{gravi, st1983}. 
This barrier is produced by the centrifugal force and the 
height of this barrier in general relativistic calculation depends on the
magnitude of the angular momentum of the particle. Depending on the energy
content of the particle, this potential barrier
can give rise to a capture orbit or deflect particle away
from the black hole or even allow the particle to be accreted by
the black hole \citep{st1983}. Though the fluid motion
is fundamentally different from a particle motion, such general
conclusions derived from particle dynamics also hold for
fluid motion. Thus, for the studies of accretion onto black
holes, many research groups
prefer to solve general relativistic fluid dynamic equations.

Over the past 50 years, several numerical simulation codes
for solving the time-dependent general relativistic fluid dynamics
equations have been developed 
\citep[][and references therein]{wilson1972,hawley1984,Banyuls1997a,del2002,harm,font2008,bhac,porth2019}.
Many of these simulation frameworks assume the total mass of the 
accretion disk to be negligible compared to the
central accretor so that the space-time metric remains unchanged
throughout the simulation. Current practice in the
accretion disk simulation community is to use an equilibrium thick
disk solution threaded by a seed magnetic field as the initial
condition \citep{abra2013}. 
This seed magnetic field generates some instability inside the disk
so that the equilibrium condition is broken and a significant
part of matter starts getting accreted onto the black hole.

In contrast to this, one can consider a more realistic set up
where the matter comes from, ideally, an infinite distance
(e.g, the companion star in the case of BHXRB or ISM in the case of black hole
at the center of galaxy) to the accreting black hole.
Due to limited computational resources, we of course
cannot simulate the entire accretion process.
So, we simulate the dynamically important inner part of the
disk and rely on the analytical solution for the rest of the part.
To simulate such a configuration, one considers the matter to enter the
simulation domain through the outer boundary located at a finite
distance from the black hole. And, inside the simulation domain,
the solution of the time-dependent equations self-consistently
determines the flow configuration. Thus, one applies an appropriate inflow
boundary condition at the outer boundary and supplies the flow
parameters there. These inflow parameters are chosen in such a way that
they are consistent with certain analytical solution that
extends up to the infinite distance.
Several analytical advective accretion solutions are present in 
literature which connects black horizon with infinite distance
(e.g., Bondi accretion solution \citep{bondi1952}, 
slim disk \citep{abra1988}, hybrid model flows \citep{Chakrabarti1989a},
ADAF solution \citep{narayan1994} etc.).
Therefore, at the outer boundary of the simulation domain one can use
the analytically calculated values of flow parameters using such
solutions.

Simulations of purely radial, Bondi type accretion flow does not produce
significant variability. Rather, simulations of quasi-spherical,
rotating accretion flow show
interesting flow features. Accretion disk simulations with 
Bondi-type density and radial velocity distribution along with 
an additional, arbitrary latitude-dependent
angular momentum component as the inflow boundary condition
have been conducted using non-GR \citep{proga2003,janiuk2008,li2013}
as well as GR codes \citep{sukova2017,ressler2021,lalakos2022,kaaz2023,
cho2023,olivares2023,dihingia2024}.
We are rather interested in solutions of rotating flows
which self-consistently takes the matter rotation into account.  
Steady state solutions, extending from horizon to infinity,
of the rotating fluids with sub-Keplerian
angular momentum (at a large distance from a black hole)
are present in literature \citep{fukue1987, chakraba1989}.
The solution shows that the centrifugal barrier can slow
down the nearly free fall motion of the flow close to the black hole. 
This barrier can even force the fluid to pass
through shocks before being accreted.
These solutions are advective and show significantly high value
of radial velocity except near the shock location where the flow
is nearly halted. When the angular momentum is set to zero,
the solution becomes identical to the standard Bondi accretion
solution.

Numerical simulations of this flow in the close vicinity
of the black holes have resulted in a dynamical flow configuration
which can explain several observational features 
\citep{Molteni1994a,rcm1997,Chakrabarti2004a,Garain2014a,Das2014a,
Lee2016a,patra2019,debnath2024}.
Many of these simulation works
are done using pseudo-Newtonian potential proposed in \citet{Paczynsky1980a}.
This potential reproduces the effective potential
around a non-rotating black hole satisfactorily well.
However, being non-general relativistic, it has several drawbacks
compared to the Schwarzschild spacetime. For example, computations
show that the fluid bulk velocity can become super-luminal close
to the horizon. Additionally, the effects of space-time dragging
due to black hole rotation cannot be investigated. For these
reasons, it is beneficial to study such accretion flow using
a general relativistic fluid solver in the Kerr background.
A few simulation results for this kind of flow with above-mentioned
set-up are reported in \citet{kgbc2017,kgcb2019}. 
However, further extensions are not reported.
In this paper, we describe the development of a general relativistic 
hydrodynamics simulation code, designed specifically to implement
the above set-up and study such advective flow. We provide results of
several validating test problems where we compare the numerical solutions 
obtained using this simulation code with the analytical solutions
for the above mentioned advective flow.

Our paper is organized as follows: In Section~\ref{sec:1.5},
we provide a very brief overview of the general relativistic
analytical solution of the sub-Keplerian accretion flow. We shall
use these solutions as our benchmark test problems.
In Section~\ref{sec:2},
we introduce the time-dependent general relativistic 
hydrodynamics (GRHD) equations and our numerical solution
methodology. In Section~\ref{sec:3}, we present the results.
Finally, in Section~\ref{sec:4}, we provide a summary and our
concluding remarks.

In our following calculations, we use $r_g=GM_{bh}/c^2$ as unit
of distance, $r_g/c$ as unit of time and $r_gc$ as unit of 
specific (i.e., per unit mass) angular momentum. 
Specific energy is measured in the unit of $c^2$.
Here, $G$ is the gravitational constant, $M_{bh}$ is the
mass of the black hole and $c$ is the speed of light in vacuum.

\section{Theory of Sub-Keplerian advective flow}
\label{sec:1.5}
General relativistic, steady state sub-Keplerian 
advective flow solution is discussed in great
details in many references \citep{Chakrabarti1990b,Chakrabarti1996d,
Chakrabarti1996b,Chakrabarti1996c}. Here, we provide a very brief
discussion for the sake of completeness. Analytical solution aims to find
the radial variation of solution variables under various flow models
such as wedge flow, constant height flow and vertical equilibrium.
For analytical studies, we use the following form of Kerr 
metric (valid only near the equatorial plane) expressed in 
cylindrical coordinates:
\begin{equation}
\label{eqgmunu}
\begin{split}
ds^2&= g_{\mu\nu}dx^{\mu}dx^{\nu}\\
&= - \frac{r^2\Delta}{\delta}dt^2 + \frac{\delta}{r^2}\left( d\phi - \omega dt\right)^2
+ \frac{r^2}{\Delta}dr^2 + dz^2
\end{split}
\end{equation}
\citep{Novikov1973a}. Here,
$$
\delta=r^4+r^2a^2+2ra^2,~\Delta=r^2-2r+a^2,~\omega=2ar/\delta, 
$$
$a$ being the spin parameter of the black hole.

Steady state solution
of non-dissipative sub-Keplerian advective flow is derived
using the conservation equations of mass accretion rate
$\dot{m}$ and specific energy $\epsilon$:
\begin{eqnarray}
\dot{m} =  \rho  u^r A\\
\epsilon = h u_t = \frac{1}{1 - na_s^2} u_t.
\label{eq:01}
\end{eqnarray}
Here, $\rho$ is the rest-mass density, $u^r$ is $r$ component of the
four-velocity $u^\mu$, $A$ is a geometric quantity representing 
the surface area through which mass flux is considered. For different
flow models, $A$ may have different expressions. $h=1/(1-na_s^2)$
represents the enthalpy with $a_s$ being the sound speed. $u_t$ is 
the $t$ component of the four velocity $u_\mu$ and is obtained using
the normalization $u^\mu u_\mu = -1$ as follows:
\begin{equation}
u_t=\left[\frac{\Delta}{(1-V^2)(1-\Omega l)(g_{\phi\phi}+l g_{t\phi})}\right]^{1/2}.
\end{equation}
Here, $\Omega$ is the angular velocity of the rotating fluid
\begin{equation}
\label{omega}
\Omega = \frac{u^\phi}{u^t}=-\frac{g_{t\phi}+lg_{tt}}{g_{\phi\phi}+lg_{t\phi}},
\end{equation} 
and $l=-u_\phi/u_t$ is the specific angular momentum (angular momentum
per unit mass) and it is a conserved quantity.
Also, the radial velocity $\mathcal{V}$ in the rotating frame is 
\begin{equation}
\label{bigV}
\mathcal{V}=\frac{v}{\left(1-\Omega l\right)^{1/2}},
\end{equation} 
where 
\begin{equation}
\label{smallv}
v=\left(-\frac{u_ru^r}{u_tu^t}\right)^{1/2}.
\end{equation}

Next, one takes derivative of $\dot{m}$ and $\epsilon$ w.r.t. $r$
and eliminates $da_s/dr$ from both the equations and finally, obtains an
equation for $d\mathcal{V}/dr$. Imposition of transonic condition in this
equation enables us to find the critical (transonic) points.
This imposition implies that such a transonic solution depends only
on two conserved parameters out of three, namely, $\dot{m}$,
$\epsilon$ and $l$ \citep{Chakrabarti1990b}. 

\begin{figure}[h!]
\begin{center}
\includegraphics[scale=.49]{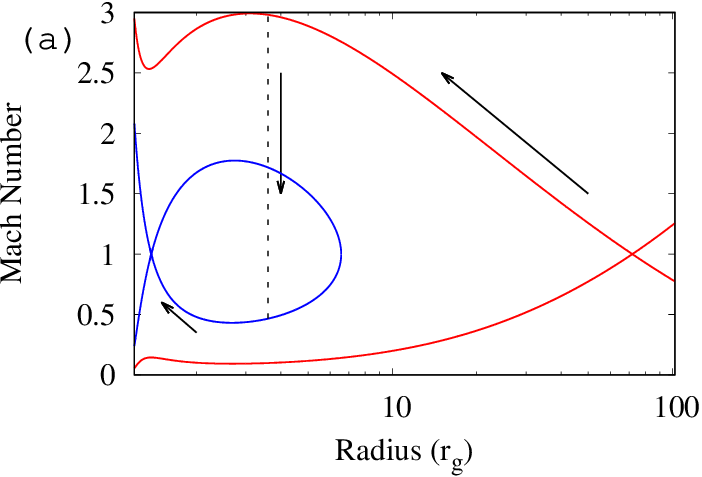}
\includegraphics[scale=.49]{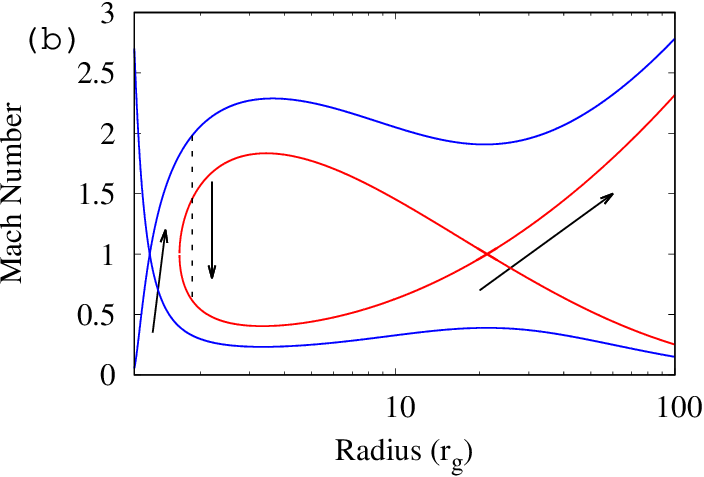}
\end{center}
%
%
\caption{
a) shows the radial variation of Mach number $\mathcal{V}/a_s$
for an accretion solution,
whereas, b) shows the same for a wind solution. In both the Figures,
the arrows indicate the solution branches followed by the flow. See text
for details.
        }
\label{fig:0}       
\end{figure}

The equation of $d\mathcal{V}/dr$ is numerically solved to find
$\mathcal{V}(r)$. For certain combinations of flow parameters $\epsilon$ and $l$,
the solutions show presence of two X-type critical points. For
accretion or wind solution, a solution branch passing through the
outer critical point and a solution branch passing through
the inner critical point may be connected via a shock jump.
The shock location is found by applying Rankine-Hugoniot condition.
Fig. \ref{fig:0} shows radial variation of Mach number $\mathcal{V}/a_s$ for the
accretion and the wind solutions for a Kerr black hole with spin
parameter $a=0.99$.
Fig. \ref{fig:0}(a) shows the example of the accretion solution.
This is done for $\epsilon=1.01$ and $l=2.073$. 
Red solid line shows the solution branches passing through the
outer critical point located at $r=71.84$
and the blue solid line shows the same passing through the
inner critical point located at $r=1.38$. The shock location
at $r=3.6$ is shown by the dashed line. The arrows indicate
the solution branches followed by the accreting matter.
Similarly, Fig. \ref{fig:0}(b) shows an example of the wind solution.
This is done for $\epsilon=1.03$ and $l=2.1$. Line styles are
same as in Fig. \ref{fig:0}(a). The outer critical point, the inner critical
point and the shock locations are 21.26, 1.32 and 1.87 respectively.
We are going to use these analytical solutions for benchmarking
our code.

\section{GRHD: Basic equations and solution procedure}
\label{sec:2}
GRHD equations are derived from
the following conservation laws:
\begin{eqnarray}
\nabla_\mu \left(\rho u^\mu\right) = 0 \\
\nabla_\mu T^{\mu\nu} = 0
\label{eq:01}
\end{eqnarray}
Here, $\nabla_\mu$ represents the covariant derivative,
$u^\mu$ is the four-velocity and 
$T^{\mu\nu}$ is the stress-energy tensor. 
$T^{\mu\nu}=\rho h u^\mu u^\nu + P g^{\mu\nu}$ for ideal fluid
with $h$ as the specific
enthalpy given by $h=1+\frac{\Gamma}{\Gamma -1}\frac{P}{\rho}$, 
$\Gamma=4/3$ being the adiabatic index and $P$ being the pressure.
The first equation represents the conservation of baryon number
and the second equation is the conservation of energy-momentum
tensor.

Following \{3+1\} formalism \citep{Banyuls1997a,font2008}, 
we write the space-time metric 
$g_{\mu\nu}$ in terms of lapse ($\alpha$), shift vector
($\beta^i$) and the spatial metric ($\gamma_{ij}$).
After some algebraic manipulations, 
this set of equations can be written as a set of five
partial differential equations (PDEs):
\begin{eqnarray}
\frac{1}{\sqrt{-g}}\left[\frac{\partial \sqrt{\gamma} D}{\partial t} +
		      \frac{\partial}{\partial x^i}
		      \left(\sqrt{-g}D\left( v^i - \frac{\beta^i}{\alpha}\right)\right)\right]&=&0 \\
\frac{1}{\sqrt{-g}}\left[\frac{\partial \sqrt{\gamma} S_j}{\partial t} +
		      \frac{\partial}{\partial x^i}
		      \left(\sqrt{-g}\left(S_j\left( v^i - \frac{\beta^i}{\alpha}\right)+P\delta^i_j\right)\right)\right]&=&
		      T^{\mu\nu}\left(\frac{\partial g_{\nu j}}{\partial x^\mu}-\Gamma^\lambda_{\nu \mu} g_{\lambda j}\right) \\
\frac{1}{\sqrt{-g}}\left[\frac{\partial \sqrt{\gamma} \tau}{\partial t} +
		      \frac{\partial}{\partial x^i}
		      \left(\sqrt{-g}\left(\tau\left( v^i - \frac{\beta^i}{\alpha}\right)+P v^i\right)\right)\right]&=&
		      \alpha \left(T^{\mu 0}\frac{\partial \mathrm{ln}\alpha}{\partial x^\mu} - T^{\mu\nu}\Gamma^0_{\mu\nu}\right)
\end{eqnarray}

Here, $\sqrt{-g}\equiv det(g_{\mu\nu})$ and 
$\sqrt{\gamma}\equiv det(\gamma_{ij})$,
and these are connected by $\sqrt{-g}=\alpha \sqrt{\gamma}$. We denote
the set of five-component vector $U=\left(D, S_j, \tau\right)$ 
as vector of conserved 
variables and can be expressed in terms of vector of primitive
variables $V=\left(\rho, v^i, P\right)$ as follows:
$$
D=\rho W, \quad S_j = \rho h W^2 v_j, \quad \tau = \rho h W^2 - P - D.
$$
Here, $W$ is the Lorentz factor given by 
$W=1/\sqrt{1-v^iv_i}=\alpha u^t$.
$v^i$ are the components of three-velocity given as
$v^i=\frac{u^i}{\alpha u^t} + \frac{\beta^i}{\alpha}$ and the 
co-variant counterpart can be calculated as $v_j=\gamma_{ij}v^i$.

The above set of PDEs is further written in integral form and
subsequently discretised on a given mesh \citep{Banyuls1997a,font2008}. 
The resulting discretised equations
on a spherical mesh constructed using Boyer–Lindquist 
coordinates $(t,r,\theta,\phi)$ \citep{Boyer1967a} 
are solved using finite volume method.
For our present calculations,
we use the following form of the Kerr space-time metric:
\[
g_{\mu\nu} =
\begin{bmatrix}
-(1-\frac{2r}{\sigma}) & 0 & 0 & -\frac{2ar \sin^2\theta}{\sigma} \\
	0             & \frac{\sigma}{\Sigma} & 0 & 0 \\
	0             & 0         &  \sigma   &  0 \\
-\frac{2ar \sin^2\theta}{\sigma} & 0 & 0 & ( r^2 + a^2 + \frac{2ra^2 \sin^2\theta}{\sigma})\sin^2\theta
\end{bmatrix},
\]
where, $\sigma = r^2+a^2\cos^2\theta$ and $\Sigma=r^2 - 2r + a^2$.
With these notations, the lapse $\alpha$ and shift $\beta^i$ functions
are as follows:
$$
\alpha=\sqrt{\frac{\sigma \Sigma}{(r^2 + a^2)\sigma + 2ra^2 \sin^2\theta}},
$$
$$
\beta^r=\beta^\theta=0,\quad \beta^\phi =  -\frac{2ar}{(r^2 + a^2)\sigma + 2ra^2 \sin^2\theta}.
$$
Also, the determinant of the metric is
$$
\sqrt{-g}=\sigma\sin\theta/\alpha
$$

This hydrodynamics simulation code is an extension of our
non-GR code used in a previous work \citep{garain2023}. We improve
several subroutines suitably to incorporate general relativistic effects.
For spatial reconstruction, we have used second order accurate
van Leer slope limiter following \citep{mignone2014}. We perform 
reconstruction on vector $\left(\rho, Wv^i, P\right)$, instead of
primitive variable vector $V$, since the reconstruction on $Wv^i$
ensures sub-luminal reconstructed profile of $v^i$ inside a zone
\citep{bk2016}. We have provisions for HLL and LLF Riemann solvers
for calculating the interfacial fluxes. For all the presented results,
we use HLL Riemann solver. Second-order accurate
strong stability preserving Runge-Kutta(RK) time integration is used
for time advancement. One of the non-trivial step in GRHD
is the conserved-to-primitive conversion
as it requires a non-linear equation solution employing a root
solver (e.g., Newton-Raphson).
We have implemented two methods following \citet{mb2005} and
\citet{del2002}. For our calculations, we prefer the method of
\citet{del2002}. It may happen that the root solver does not converge for
a few zones after the time-update step and
for such zones, we use the previous
time-step solution as it is already saved in a RK type time-update.
The timestep $dt$ is calculated following standard
Courant-Friedrichs-Lewy (CFL) condition \citep{leveque2002,toro2009}
$$
dt = C_\mathrm{CFL}\frac{1}{\frac{\lambda^r}{dr} 
                  + \frac{\lambda^\theta}{r d\theta}
		  + \frac{\lambda^\phi}{r\sin\theta d\phi}},
$$
where, $\lambda^i$ is maximum characteristic speed in 
$i^\mathrm{th}$ direction
and $C_\mathrm{CFL}$ is the CFL number. For all the runs, we use
$C_{\mathrm CFL} = 0.9$. For one- or two-dimensional simulations,
contribution from the corresponding inactive dimension(s) is switched
off.

\section{Results}
\label{sec:3}
In this section, we present results of a few standard test 
problems validating our implementation. Later in this section, we present
results of multi-dimensional simulation for sub-Keplerian advective
accretion disk.


\subsection{Accuracy analysis using two-dimensional equilibrium torus:}

Equilibrium torus is a hydrostatic equilibrium solution of the above mentioned
GRHD equations. The solution results in a geometrically thick disk
around a gravitating source and the matter is held at its position
because of the balance between the inward gravitational pull and the combined
effect of outward centrifugal and pressure gradient forces.
General procedure for the construction of such disks is given in
many papers, e.g., \citet{abra1978,Chakrabarti1985,font2002} etc. 
In this work, we consider a constant angular momentum disk and
calculate the density, pressure and velocity distributions
following the analytical calculations given in the above references.
Next, we initialize our computational domain with these distributions
and run our simulation for some time. Since this is a
time-steady solution, we expect the values of these variables to
remain same at all positions. However, due to inherent errors
of the numerical solution scheme, the numerical solution is expected to 
develop error.

We conduct several two dimensional ($r,\theta$) simulations
of such thick disk around black holes with different
spin parameters. Here, we show results for two cases, 
(1) with spin parameter $a=0$ and (2) with $a=0.99$.
In our simulations, we maintain the initial state of the solutions
at the ghost zones of both the radial boundaries throughout 
the simulations. Such fixed boundary conditions have been used
previously for global accuracy convergence demonstrations 
\citep{mignone2014,fambri2018,ced2018,weno2020}.
On the $\theta$ boundaries, we use reflection
boundary condition.

\begin{figure}[h!]
\begin{center}
\includegraphics[scale=.49]{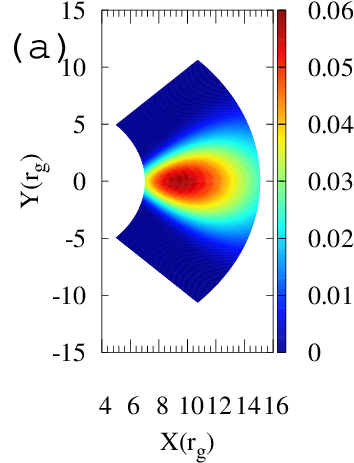}
\includegraphics[scale=.49]{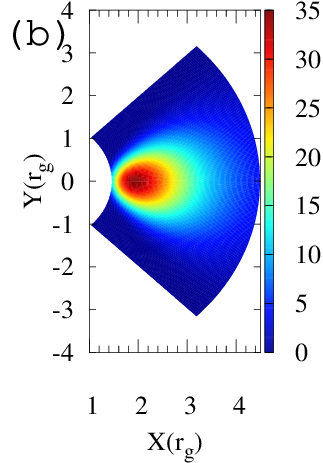}
\end{center}
%
%
\caption{
(a) shows the density distribution inside the thick disk 
at the final time $t=100$ for the case with spin $a=0$.
The disk center is located around 9.3.
(b) shows the same at $t=100$ for the case with spin $a=0.99$.
The disk center is located around 1.94.
        }
\label{fig:1}       
\end{figure}

\begin{figure}[h!]
\begin{center}
\includegraphics[scale=.49]{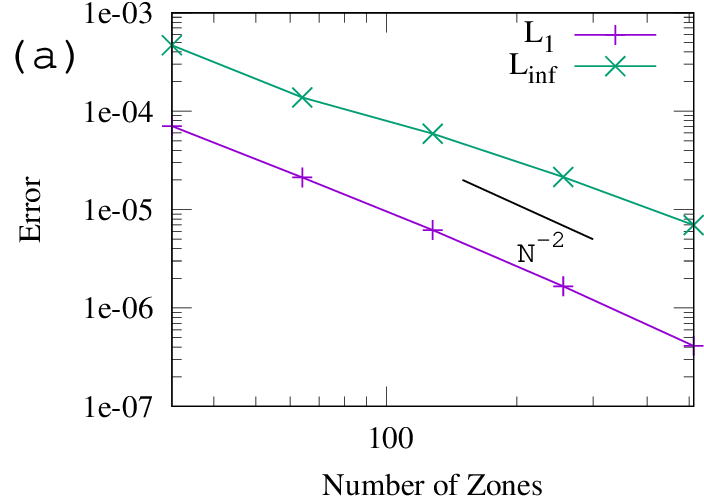}
\includegraphics[scale=.49]{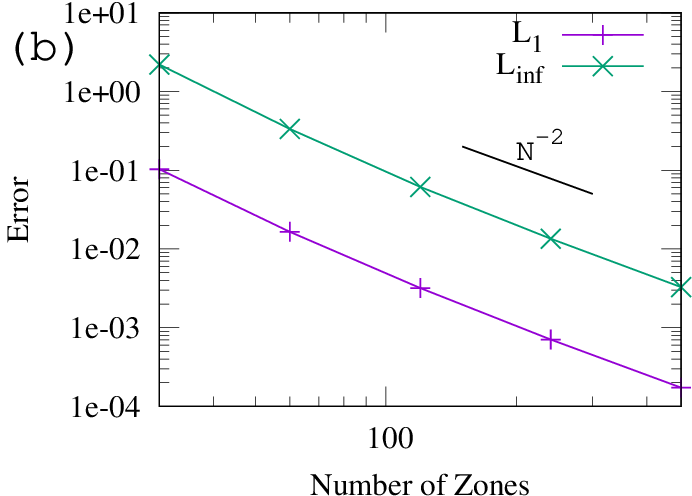}
\end{center}
%
%
\caption{ Accuracy demonstration for torus problem:
(a) shows the $L_1$ and $L_{inf}$ error convergence results
for the case with spin $a=0$.
(b) shows the same for the case with spin $a=0.99$.
        }
\label{fig:2}       
\end{figure}

For case (1), we choose the constant specific angular momentum to be
$l=3.9$. This disk has a cusp at around 4.3 and the disk center is
located around 9.3.
We use inner edge of the disk at 4.5 while calculating the fluid
variables so that the disk does not fill up the Roche lobe.
For accuracy analysis, the simulations are run on varying
grid sizes ranging from [32x45] to [512x720] on a 
$r-\theta$ domain [7:15]x[$\pi$/4:3$\pi$/4]. 
The disk center is located inside our computational domain.
Fig. \ref{fig:1}(a) shows
the density distribution for this case at the final 
time $t=100$. This result is drawn for the simulation
with grid size [128x180]. Fig. \ref{fig:2}(a)
shows the result of the accuracy
analysis in the density variable. Purple line shows the convergence
result for $L_1$ error while
green line shows the same for $L_{inf}$ error. The reference slope is
also provided for comparison.

For case (2), we choose $l=2.19$. This disk has a cusp at around
1.22 and the disk center is located around 1.94. We use inner 
edge of the disk at 1.22 for this case. The simulations are run 
in the $r,\theta$ domain [1.45:4.45]x[$\pi$/4:3$\pi$/4]
on grid cells [30x30] to [480x480]. Fig. \ref{fig:1}(b) shows the 
density distribution for this case at the final time $t=100$.
This result is drawn for the simulation with grid size [120x120].
Fig. \ref{fig:2}(b) shows the result of the accuracy
analysis in the density variable. Purple line shows the convergence
result for $L_1$ error while
green line shows the same for $L_{inf}$ error. The reference slope is
also provided for comparison.

\subsection{Shock in one-dimensional advective flow:}

Here we demonstrate our code's capability to reproduce the steady state advective
flow solutions having a shock.
First, we compare the analytical solution of a constant, low-angular
momentum (sub-Keplerian) accretion flow onto a black hole with
the numerically simulated one. Next, we perform similar comparison
for a wind solution.

Sub-Keplerian accretion
solution connects infinity to the black hole horizon. The
subsonic matter at far away (infinite) distance accelerates 
to supersonic speed at a finite
distance (sonic point) from black hole before it reaches the black hole
horizon. As discussed in Section~\ref{sec:1.5}, if the flow
specific energy ($\epsilon$) and specific angular momentum ($l$) fall
within certain parameter space, the flow may pass through a shock
after crossing the sonic point.
In such case, the matter accelerates soon to become supersonic again 
at certain radius (inner sonic point) before reaching the horizon.

For our comparison, we choose flow parameters such that the
solution contains a shock. Theoretically, solution branches passing
through the outer and the inner sonic points have different
entropies \citep{Chakrabarti1989a}: solution through the inner sonic point has
higher entropy than that through the outer one.
Thus, thermodynamically, as the matter approaches the black hole, 
it prefers to follow the solution through the inner sonic point. 
However, note from Fig. \ref{fig:0}(a) that the solution through the inner
sonic point (blue solid line) does not extend up to infinite distance.
Thus, the matter coming from infinite distance initially follows 
the solution passing through the outer sonic point. However, when the matter
arrives sufficiently close to the black hole so that solution through
inner sonic point is available, it makes a transition to this solution. 
The required excess entropy is produced at the shock and that allows matter
to jump from the branch passing through the outer sonic point
to the branch passing through the inner sonic point.
In a realistic three-dimensional flow, the matter bouncing off the centrifugal
barrier collides with the incoming matter and makes the flow turbulent.
This generates the required excess entropy \citep{Chakrabarti1993a}.

\begin{figure}[h!]
\begin{center}
\includegraphics[scale=.49]{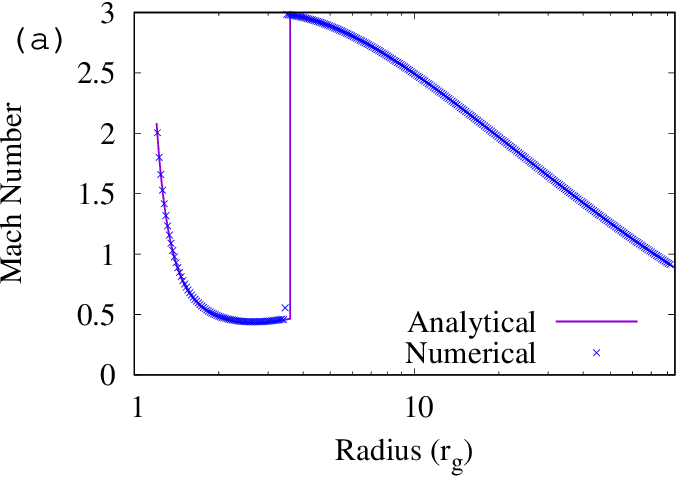}
\includegraphics[scale=.49]{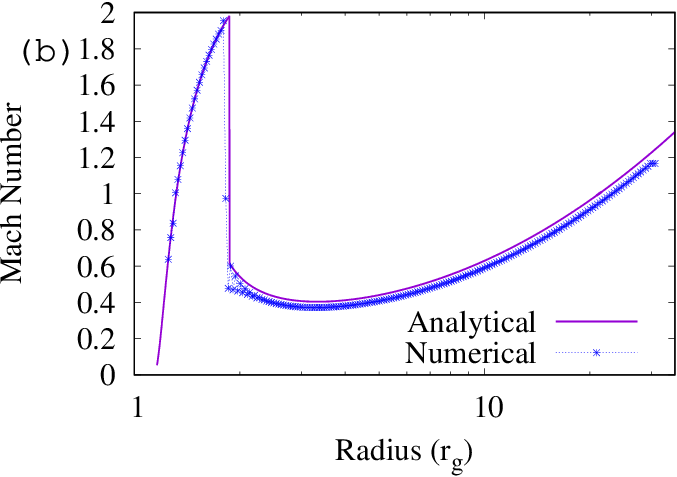}
\end{center}
%
%
\caption{
(a) shows the shock in accretion flow solution, whereas,
(b) shows the same in wind solution. See text for details.
        }
\label{fig:3}       
\end{figure}

Shocks in an accretion flow can form at a few hundreds to only a few
$r_g$ distance depending on ($\epsilon,l$) pair. Capturing
a shock very close a black hole is one of the stringent
tests.
In Fig. \ref{fig:3}(a), we show that our simulation code
can capture a shock at $r=3.5$.
Radial variation of Mach number is shown in this plot.
We see that the shock is resolved within one grid point. This solution
corresponds to $\epsilon=1.01$ and $l=2.073$.
Solid line shows the analytical solution (solution marked by arrows
in Fig. \ref{fig:0}(a)) and the crosses
are the numerically simulated solution.
This simulation is carried out using 300 logarithmically binned grid
cells
inside the domain [1.2:80]. Black hole spin is assumed to be $a=0.99$.
We use outflow boundary condition at the inner edge and
inflow boundary condition at the outer edge. We use two ghost cells  
for our computation. The centroids of these cells are located
at 80.566 and 81.702.
As the inflow boundary, we maintain $V=\left(1, 7.303E-2, 0, 3.153E-4, 4.9089E-3\right)$
at the first cell and $V=\left(0.985, 7.213E-2, 0, 3.067E-4, 4.808E-3\right)$ at
the second cell throughout the simulation. Initially, the
computation domain is filled with static matter with 
$\rho=\rho_{\rm floor}=10^{-8}$ and
$P=P_{\rm floor}=a_s^2*\rho_{\rm floor}/\gamma$ where $a_s=0.0802$ is the
sound speed at the outer radial boundary. $\gamma=4/3$
is assumed for this run. Thus, as the simulation is
started, matter rushes towards the black hole and within a few
hundred $r_g/c$ time, simulation domain is filled up with matter that
corresponds to the analytical solution passing through the outer
sonic point. Since one-dimensional flow doesn't develop turbulence,
this solution does not pass through the shock and the inner sonic point.
Thus, to produce shock in a one-dimensional flow, we need to
momentarily apply some perturbation such that the flow acquires 
sufficient additional entropy to pass through the inner sonic point
\citep{Chakrabarti1993a}.
Once the flow passes through the inner sonic point, it automatically
develops a shock which ultimately settles down close to
the theoretically predicted location.
In our simulation, this perturbation is applied
at the inflow boundary: we momentarily (for a duration of 30 $r_g/c$)
increase the pressure by a factor of 9 in the two outer radial
boundary ghost cells at around time $t=4000$.
This perturbation is advected with the flow and makes it pass through
the inner sonic point and a stable solution passing through the
shock is developed by time $t=7000$. We run the simulation
till a stopping time of $t=50000$ just to ensure that it is
actually a steady state. Analytical calculation provides the
location of outer sonic point, shock and inner sonic points at
71.84, 3.6 and 1.38, respectively. Our numerical calculation
captures the outer and inner sonic points exactly at these locations
as can be seen in Fig. \ref{fig:3}(a). The numerical calculation
finds the shock at 3.5.
 
Just like the accretion flow solution can pass through 
a shock, a wind solution can also pass through a shock
close to the black hole if the $\epsilon$ and $l$ values
for the solution are chosen within a certain range.
Fig. \ref{fig:3}(b) shows the simulation of such a wind solution. 
Radial variation of Mach number is shown in this plot. 
This solution corresponds to $\epsilon=1.03$ and $l=2.1$.
Solid line shows the analytical solution (solution marked by arrows
in Fig. \ref{fig:0}(b)) and the crosses
are the numerically simulated solution.
This simulation is carried out using 200 logarithmically binned grid
cells
inside the domain [1.28:30]. Black hole spin is taken to be $a=0.99$ for this
case. For simulating
wind solution, matter is launched from very close to the black hole
and it flies off to infinite distant.
Thus, we use inflow boundary condition at the inner radial boundary
and outflow boundary condition at the outer radial boundary.
We use two ghost cells for our computation. The centroids of the two
inner ghost cells are located at around $r=1.25$ and $r=1.27$.
As the inflow boundary, we maintain $V=\left(1.317, 0.0305, 0, 0.2725, 0.187\right)$
in the first cell and $V=\left(1, 0.0384, 0, 0.2756, 0.1297\right)$ in
the second cell throughout the simulation. 
As the initial condition, we fill up part of the simulation domain
[1.28:$r_b$] with analytical solutions passing through the inner 
sonic point and the rest [$r_b$:30] with the analytical solution
passing through the outer sonic point. Thus, the initial state
has a shock discontinuity at $r_b$. For result shown in Fig. \ref{fig:3}(b)
, we use $r_b=3$. With this initial state, we
start the simulation and within time around $t=2000$, the shock discontinuity
settles down at a radius $r=1.87$ which is very close to the
theoretically predicted shock location $=1.82$.
Note that for this case, the shock is inside the ergosphere.
For our code, we find the solution immediately after the shock is oscillatory.
Possibly, this can be avoided by using higher order spatially accurate reconstruction.
We run the simulation till a stopping time of $t=10000$ just to 
ensure that it is actually a steady state.
We have simulated different cases with different values of $r_b$ and ensure
that this steady state solution is independent of $r_b$ value.


\subsection{Two-dimensional Bondi accretion ﬂow:}

This test problem is adapted from \citet{kgcb2019}. In this problem,
we show the effect of spacetime dragging close to a rotating black
hole. Because of this dragging, a spherically symmetric accretion
configuration onto a rotating black hole develops axisymmetry
close to the black hole (see also \citet{aqua2021}).

This simulation is performed on a $r-\theta$ computation domain
[1.35:200]x[0:$\pi$] using 300x180 grid cells. In the $r$ direction,
we use logarithmic binning and in the $\theta$ direction, grids
are equispaced. Spin parameter of the black hole is assumed to
be $a=0.99$. The boundary conditions are same as in
\citet{kgcb2019}. Matter enters simulation domain with
$\mathcal{V}=0.0229$ and $a_s=0.0805$ at $r_{\rm out}=200$. This corresponds
to $\epsilon = 1.015$ and $l=0$. At the inner
radial boundary, we use zero gradient outflow boundary condition.
At the $\theta$ boundaries, we use reflecting boundary conditions.
The simulation is run till a stopping time of 20000. By this time,
a steady state solution has developed.

\begin{figure}[h!]
\begin{center}
\includegraphics[scale=.53]{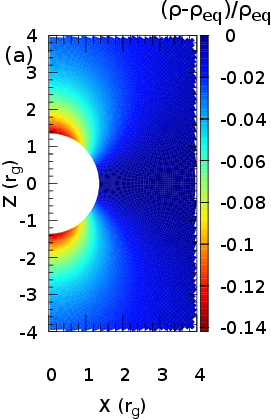}
\includegraphics[scale=.49]{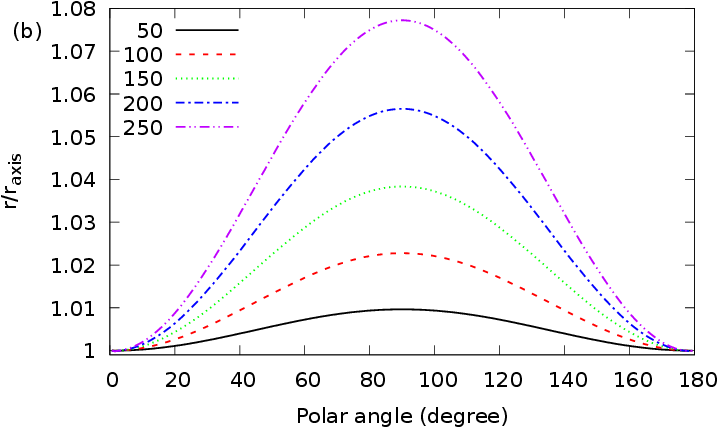}
\includegraphics[scale=.53]{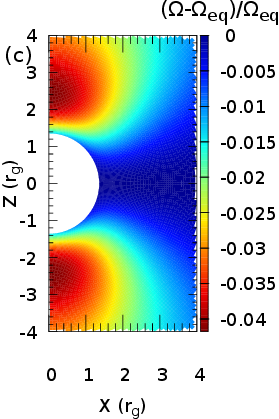}
\end{center}
%
%
\caption{
(a) shows the distribution of relative density w.r.t equatorial density
at final time $t=20000$,
(b) shows the radial coordinate $r$ normalized by $r$ on axis for a
given density value as a function of polar angle,
(c) shows the distribution of relative $\Omega$ w.r.t equatorial $\Omega$.
See text for details.
        }
\label{fig:5}       
\end{figure}

In this test problem, matter is injected into the simulation
domain spherically symmetrically. In the
absence of black hole rotation, such accretion solution remains
spherically symmetric by the time it crosses the horizon. However,
a rotating black hole drags the spacetime around it and hence breaks
the spherical symmetry and makes the accretion solution axisymmetric.
Such axisymmetry can be visualized in the fluid variables such as density
distribution. To quantify this, we plot the distribution of
$(\rho(r,\theta) - \rho_{\rm eq}(r,\pi/2))/\rho_{\rm eq}(r,\pi/2)$,
where, $\rho_{\rm eq}(r,\pi/2)$ is the density at ($r,\theta=\pi/2$),
in the inner part of the simulation domain.
Colors in Fig. \ref{fig:5}(a) show this quantity. We can clearly
see that, at a given radius $r$, $\rho$ towards the polar region
is less than $\rho$ at the equator. The maximum difference
between the density values is nearly 14\% for this simulation.
We also trace the density iso-contours on the $r-\theta$
plane in Fig. \ref{fig:5}(b). On the x-axis, we plot polar
angle $\theta$ and on the y-axis, we plot the radial coordinate
$r$ normalized by $r$ on the axis for a given density value.
Contours are drawn for density values 50, 100, 150, 200 and 250
(from bottom to top). This plot shows that same density value
appears at larger $r$ on the equator than on the poles. For
example, highest density value 250 appears at 7.8\% higher 
$r$ on the equator than on the poles. This Fig. can be compared
with Fig. 2(b) of \citet{kgcb2019} who obtained a value slightly
higher than 7\% (less than our value of 7.8\%). This is caused
by our choice of higher black hole spin $a=0.99$ compared 
to their $a=0.95$. Frame dragging induces axisymmetric rotation 
$\Omega(r,\theta)=u^\phi/u^t$ although specific angular momentum
$l=-u_\phi/u_t$ is zero for this flow. Fig. \ref{fig:5}(c) shows
the distribution of 
$(\Omega(r,\theta) - \Omega_{\rm eq}(r,\pi/2))/\Omega_{\rm eq}(r,\pi/2)$,
as in Fig. \ref{fig:5}(a). We see that $\Omega$ is lesser towards
the pole.

\subsection{Two-dimensional sub-Keplerian accretion flow:}

This is the multi-dimensional extension of the one-dimensional
advective flow discussed in Section~\ref{sec:1.5}.
In multi-dimensional simulations, one gets an opportunity
to study the vertical structure of the geometrically
thick sub-Keplerian advective flow. The simulation procedure is nearly
same as the above mentioned two-dimensional Bondi accretion flow. Only
difference is at the implementation of the outer radial 
boundary condition.
Instead of spherically symmetric inflow at the outer boundary,
we now inject matter axi-symmetrically. Additionally, to allow
outflow from the accretion disk through the outer boundary, 
we restrict matter injection
within $-10^\circ \leq (\theta-90^\circ) \leq 10^\circ$ 
and apply zero gradient
outflow boundary condition otherwise.

This simulation has been performed on a $r-\theta$ 
domain [1.35:100]x[0:$\pi$] using 200x180 grid cells.
In the radial direction, we use logarithmic binning and in the
$\theta$ direction, we use uniform mesh. Black hole spin is
assumed to be $a=0.99$. Matter enters simulation domain with
$\mathcal{V}=0.0666$ and $a_s=0.0647$ at $r_{\rm out}=100$. This corresponds
to $\epsilon = 1.005$ and $l=2.05$. This simulation has been run
till a stopping time of 20000. Solution has reached a time-steady
state by this time.

\begin{figure}[h!]
\begin{center}
\includegraphics[scale=.99]{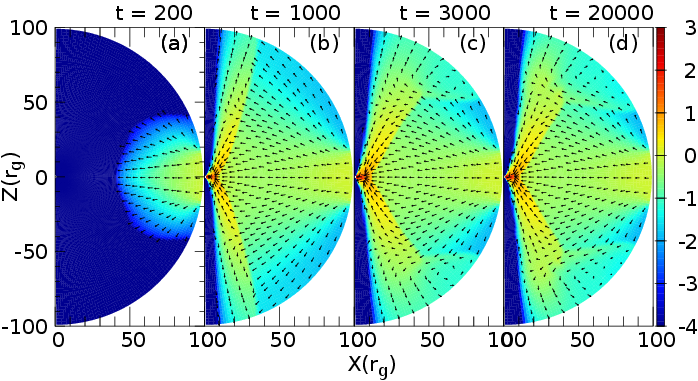}
\end{center}
%
%
\caption{
Time-evolution of density distribution on a logarithmic scale
for sub-Keplerian accretion disk simulation. Arrows show the velocity field.
        }
\label{fig:6}       
\end{figure}

Fig. \ref{fig:6} shows the sequence of snapshots at progressing time.
Colors show $log_{10}\rho$ and arrows show the direction
of velocity vectors ($v^r,v^\theta$). Length of an arrow is
proportional to the logarithm of its magnitude. Timestep is
marked on top of each Figure. Fig. \ref{fig:6}(a)
shows a transient state when the matter rushes towards the black
hole sitting at the origin through nearly vacuum. Fig. \ref{fig:6}(b),
again a transient state, shows the building of centrifugal
force supported boundary layer. The boundary layer can be identified
by tracing the density jump as we move vertically away from the 
equator. Initially, this boundary layer
expands in the radial direction and finally settles down.
By the time the solution reaches the state of Fig. \ref{fig:6}(c),
solutions becomes steady and this state continues till the
end of simulation in Fig. \ref{fig:6}(d).
The evolution is very much consistent with the earlier simulations
with non-rotating black holes (e.g.,\citet{Kim2017a}) as well as
rotating black holes (e.g., \citet{kgcb2019}) with lower spin 
($a=0.95$) than the present one ($a=0.99$).

\begin{figure}[h!]
\begin{center}
\includegraphics[scale=.99]{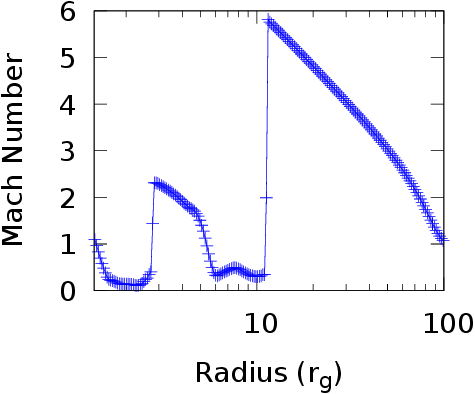}
\end{center}
%
%
\caption{
Radial variation of the Mach number along the equator
at the end of the sub-Keplerian accretion flow simulation.
        }
\label{fig:7}       
\end{figure}

Fig. \ref{fig:7} shows the radial variation of the vertically
averaged Mach number at the final time. 
Simple averaging of the Mach number values
has been done over 4 grid cells above and below the equator. Clearly,
the accretion flow develops a shock (supersonic to subsonic
transition) around $r=10$. Additionally, we find another
shock between $r=2-3$ for this case. Presence of such an inner
shock in multi-dimensional simulations has been reported earlier
for non-rotating black holes \citep{Giri2010a,Lee2011a} where
simulations are carried out using pseudo-Newtonian \citet{Paczynsky1980a}
potential. However, for a rotating black hole and using a truly 
GRHD simulation, we find such an inner shock extremely close to the horizon.
Though, as reported in \citet{cruz2012}, placement of inner boundary outside
of event horizon may affect the flow dynamics and the inner shock formation.
More elaborated analysis of such inner shocks and their
observational consequences will be discussed in future works.

\section{Summary and Conclusions}
\label{sec:4}

In this paper, we present a general relativistic hydrodynamics (GRHD)
solver. Our aim is to use the
said solver for simulating an advective accretion disk configuration
that mimics mass inflow from far out rather than starting 
from an equilibrium torus. In this solver, we solve the
GRHD equations using finite volume method on a discretised mesh
inside a given computational domain. This method incorporates
high resolution shock capturing schemes.

We have demonstrated that our presently developed GRHD 
code works for Kerr spacetime, is globally second 
order accurate and performs robustly in multi-dimensions.
For demonstrating global accuracy, we compute the numerical
errors using a hydrostatic equilibrium, geometrically thick
disk configuration and find the errors converge with second
order accuracy. We also demonstrate that our scheme can
correctly capture the analytically predicted accretion and 
wind shock solutions around a rotating black hole. 
In both the cases, shocks are resolved within one or two
grid points. Specifically, we demonstrate that our solver can
capture the shocks extremely close to the black hole (even
inside the ergosphere). Next, we demonstrate the effects of spacetime
dragging in the close vicinity of an extremely rotating black
hole (with spin parameter $a=0.99$). We show that a spherically
symmetric accretion becomes axi-symmetric as the matter
approaches the black hole. Finally, we show an example
where we simulate a geometrically thick sub-Keplerian
accretion disk. We allow rotating matter to enter the simulation
domain through a part of the outer radial boundary (close to
the equatorial region) and also allow outflow through
rest of the outer radial boundary (close to the poles).
The accreting matter rushes to the black hole through nearly 
vacuum and self-consistently forms a shock due
to centrifugal barrier. The post-shock matter forms
a thick disk. The present solution shows formation of two shocks
close to the black hole. In subsequent works, we shall explore
these thick disk solutions in great details and investigate
their radiative properties.
%
%
%
\section{Acknowledgments}
We acknowledge the usage of Kepler cluster of DPS, IISER Kolkata
and Pegasus cluster of IUCAA, Pune for running a few simulations.
Author also acknowledges the help of Mr. Pranayjit Dey during the
initial days of the code development.
%
%

\bibliographystyle{elsarticle-harv} 
\bibliography{references}






\end{document}